# Electron spin-orbit splitting in InGaAs/InP quantum well studied by means of the weak antilocalization and spin-zero effects in tilted magnetic fields.

S. A. Studenikin[a], P. T. Coleridge, G. Yu, and P. J. Poole
Institute for Microstructural Sciences, National Research Council, Ottawa, Ontario, Canada K1A 0R6

## Abstract
The coupling between Zeeman spin splitting and Rashba spin-orbit terms has been studied experimentally in a gated InGaAs/InP quantum well structure by means of simultaneous measurements of the weak antilocalization (WAL) effect and beating in the Shubnikov –de Haas (SdH) oscillations. The strength of the Zeeman splitting was regulated by tilting the magnetic field with the spin-zeros in the SdH oscillations, which are not always present, being enhanced by the tilt. In tilted fields the spin-orbit and Zeeman splittings are not additive, and a simple expression is given for the energy levels. The Rashba parameter and the electron g-factor were extracted from the position of the spin zeros in tilted fields. A good agreement is obtained for the spin-orbit coupling strength from the spin-zeros and WAL measurements.

## Introduction
The recently revived interest in the spin-orbit phenomena is driven, in part, by the subject of spintronics and by quantum information proposals that require the manipulation of spins[1,2]. In semiconductors the strength of the spin-orbit coupling can be tuned using electrostatic gates which means spins can be controlled without applying pulsed, local magnetic fields. In contrast to using magnetic fields, devices based on spin-orbit effects should be much faster, readily reduced in size and, in principle at least, integratable with other solid-state semiconductor devices.

To determine the strength of the spin-orbit coupling in two-dimensional semiconductor systems two transport techniques are commonly used[b]: The weak antilocalization (WAL) effect [3,4,5,6,7,8,9] that is seen in the very low field magneto-resistance when the spin-orbit scattering time becomes shorter than the orbital dephasing time, and the beat structure that appears in the Shubnikov-de Haas (SdH) oscillations[10,11,12] when the spin orbit interaction causes the total spin-splitting to no longer be strictly proportional to magnetic field. These two techniques are to some extent complementary. The WAL effect, a consequence of corrections associated with the quantum interference between time-reversal paths, is most apparent in low mobility samples where the magnitude of the conductivity correction (of order $e^2/\hbar$) becomes a significant fraction of the total conductivity. The beat structure is seen most clearly in high mobility samples, where the disorder is low, and where the SdH oscillations persist to lower magnetic fields.

---

[a] e-mail address: sergei.studenikin@nrc.ca
[b] In addition to transport methods there are also optical means, such as Raman and far infrared spectroscopy, allowing us to extract spin-orbit strength, e.g. see refs.[1,2] and references therein.

In some samples, such as that discussed here, the situation is complicated by a mobility that is both too large to allow an easy interpretation of the WAL effect (because the transport lifetime is larger than the spin-orbit scattering time and the standard theories[3,9] are inapplicable) and too small, so the disorder is still large enough that the SdH beat pattern is not clearly visible in small magnetic field where the spin-zero beats are expected. Even when only a single node in the beat pattern is visible, which is in principle sufficient to determine the magnitude of the spin-orbit term, there is still some uncertainty about whether this should be associated with the spin-orbit interaction or whether it is a spurious effect, associated for example with possible density inhomogeneities[13]. In such cases it is important to have an independent experiment to confirm the validity of the spin-zero and the strength of the spin-orbit interaction. We compare here results obtained in a sample where a pronounced WAL peak and an SdH spin-zero could simultaneously be observed. Hitherto, to the best of our knowledge, such a comparison made on the same sample has not been published in the literature. We show that for densities where results could be obtained using both methods they are in good agreement. This comparison was possible because (a) we have used tilted magnetic fields to enhance the strength of the single observable spin-zero and (b) in the absence of a convenient theoretical model we have proposed an empirical procedure to extract the spin orbit term from the WAL results for the situation of a transport lifetime larger than the spin-orbit scattering time. It has been assumed, as is found experimentally in similar samples[10], that the spin-orbit coupling is dominated here by the Rashba term[14] ( i.e. it is due to quantum well asymmetry) and that the contribution from the Dresselhaus term (associated with bulk inversion asymmetry[15]) is small.

## Sample characteristics

The results were obtained in a gated, modulation doped, $In_xGa_{1-x}As/InP$ quantum well sample, lattice matched to InP. The layer structure counting from the InP substrate was the following: 350 nm InP undoped buffer layer, 10 nm $In_xGa_{1-x}As$ (x=0.53) quantum well (QW), 39 nm undoped InP spacer, 13 nm *n*-doped ($N^+=4\times10^{17}cm^{-3}$) layer, 13 nm undoped InP layer, 40 nm $SiO_2$, and Au gate. Background concentration of donors in InP undoped material was $N_d\approx10^{15}$ $cm^{-3}$. Over the range of electron densities regulated by gate (from 2.5 to $4.7\times10^{15}$ $m^{-2}$) the mobility of the two-dimensional electron gas varied between 6 and 10 $m^2/Vs$. Measurements of the temperature dependence of the SdH oscillations gave an effective mass of $0.044m_e$. A measurement of the quantum mobility at $V_g=0$ ($n=4.0\times10^{15}$ $m^{-2}$) gave a value of 0.9 $m^2/Vs$, a factor of 10 smaller than the transport mobility. Such a ratio is typical for samples like this with predominant Coulomb, small angle scattering mechanism.[16]

With the magnetic field normal to the surface of the sample the SdH oscillations could typically be observed in fields down to about 0.3 Tesla with a spin-zero beat visible only for densities above $4.0\times10^{15}m^{-2}$. As the density is reduced the spin zero moves to lower fields and vanishes. However, tilting the magnetic field enhances the Zeeman component and therefore total splitting and moves the spin-zero back to higher fields, where it can again become visible. In addition, tilting the field helps to establish that the observed beat is indeed due to the spin-orbit coupling and not a spurious effect.

It was found in the sample studied that the WAL effect was prominent for all densities.[8] This implies that the spin-orbit coupling is strong, resulting in the spin-orbit scattering time $\tau_{so}$ becoming small compared to the orbital dephasing time $\tau_{\varphi}$. However, as noted above, the relatively high mobility of the sample leads to a situation where $\tau_{so}$ is also comparable with the transport scattering time $\tau_{tr}$ and where the standard theories[3,9] for an analysis of the WAL induced by the Rashba effect[4] are no longer valid. As discussed below, we have therefore had to suggest a qualitative approach for analyzing the WAL results.

## Theoretical treatment of the spin-zeroes in tilted magnetic fields

In zero magnetic field the energy splitting associated with the Rashba spin-orbit term is $\Delta_R=2\alpha k_F$, where $k_F$ is the absolute value of the Fermi *k*-vector and $\alpha$ is the Rashba coefficient.[c] In a magnetic field the spin splitting is determined by interplay between both the spin-orbit and Zeeman terms, being not just a sum of these two. The theoretical description of the spin splitting presented here follows closely the work of Das *et al.*[10] It should be noted that non-parabolicity effects are small because $n<5\times10^{11}$ cm$^{-2}$ in our experiments. The non-parabolicity corrections become noticeable (>10%) for concentrations above $10^{12}$ cm$^{-2}$.[17,18]

In the absence of a spin-orbit term the Landau level spacing $\hbar\omega_c=e\hbar B_\perp/m^*$ (with m* being the effective mass) depends only on the perpendicular component of the magnetic field, $B_\perp$, assumed to be directed along the *z* direction. In contrast, the Zeeman splitting term $g^*\mu_B B_{\text{tot}}$ (where $g^*$ is the effective electron Landé *g*-factor[d]) depends on the total applied field $B_{tot}$.

In pure perpendicular fields ($B_\perp=B_{tot}$) the total spin splitting $\delta_s$ is then given by[10]:

$$\left|\delta_s\right| = \sqrt{(1-2\beta_z)^2(\hbar\omega_c)^2+\Delta_R^2}-\hbar\omega_c \qquad (1)$$

where $\beta_z=\dfrac{g^*\mu_B B_\perp}{2\hbar\omega_c}=\dfrac{g^*m^*\mu_B}{2\hbar e}$.

When the magnetic field is tilted by an angle $\theta$ so $B_{\parallel}=B_{tot}\sin\theta$ ($B_{\parallel}$ is the parallel component of the field), the Zeeman splitting in the absence of a spin-orbit term can be written as $g^*\mu_B B_{\text{tot}}=2\sqrt{\beta_z^2+\beta_x^2}\,\hbar\omega_c$, with $\beta_x=\dfrac{g^*m^*\mu_B B_{\parallel}}{2\hbar e B_\perp}=\beta_z\tan\theta$.

The Zeeman term couples spin levels with the same Landau level number, and the Rashba term couples spin levels belonging to different Landau levels as is shown schematically in Fig1 (a). In tilted fields the effect of both terms are described by a Hamiltonian presented by the tri-diagonal matrix[10] shown in Fig. 1(b) where the diagonal, Landau level, terms $E^{\pm}_n$ also include the perpendicular part of the Zeeman splitting $\pm\beta_z$.

---

[c] Note that in literature, $\Delta_R$ is sometimes defined as half this quantity.

[d] For InGaAs, as in most other III-V compounds, the g-factor is expected to be negative.

This matrix equation must be solved numerically. Figure 2 shows (solid points) an example of such a solution for different tilts of the magnetic field. Following the procedure used by Das *et al.*[10] the size of the matrix has been chosen sufficiently large that further increases do not affect the results. The value of $\beta_z$ = -0.035 is close to the expected value for the sample studied here.[d] Positions of spin-zeroes are defined by line $\delta_s/\omega_c$=0.5 which correspond to equally spaced spin resolved levels.
While obtaining the numerical solution is a relatively straightforward procedure, it is not very convenient for fitting experimental data and extracting values for $\Delta_R$ and the electron *g*-factor. A good approximation to these results, especially for small values of $\delta_s$, can be obtained by considering only the four Landau levels closest to the Fermi energy, therefore, reducing the infinite matrix in Fig. 1 (b) to the fourth rank matrix. The resulting quadratic equation can then be solved analytically giving:

$$(\delta_s/\hbar\omega_c)^2 = P - 2Q^{\frac{1}{2}}, \text{ where}$$

$$\mathrm{P} = 2(1-\beta_z)^2 + 2(\beta_z^2+\beta_x^2) + (\Delta_R/\hbar\omega_c)^2, \qquad (2)$$

$$Q = \left[(1-\beta_z)^2 - (\beta_z^2+\beta_x^2)\right]^2 + (\Delta_R/\hbar\omega_c)^2(1+\beta_x^2)$$

Results obtained using Eq. (2) are plotted in Fig. 2 by thin solid lines. As can be seen they are generally in good agreement with the exact results (solid points) with errors of typically less than 1% for values of $\Delta_R/\hbar\omega_c$ less than about 1.5. The approximation only fails near the level crossing point of spin terms belonging to different Landau levels that occurs when $\Delta_R/\hbar\omega_c \approx 1.6$ in Fig. 2. The next spin zero occurs at higher values of $\Delta_R/\hbar\omega_c$ around 2.2, corresponding to smaller magnetic field (the spin-orbit term has a fixed value).

In the absence of a spin-orbit term the ratio of the Zeeman splitting to the cyclotron splitting can be increased without limit by tilting the field. The spin resolved levels can then cross (become evenly spaced) multiple times as the tilt angle is increased. In general, for spin resolved levels without spin orbit interaction, the amplitude of the fundamental frequency of SdH oscillations acquires a spin dependent part[19], $\cos(\pi\delta_s/\hbar\omega_c)$ which vanishes for $\delta_s/\hbar\omega_c = n + 1/2$ with *n* integer. In the absence of a spin orbit term, the spin splitting $\delta_s$ is proportional to the total field and for appropriate tilt angles the amplitude of the first harmonic of SdH oscillations vanishes over extended field range. When, however, the spin-orbit term is introduced, spin splitting $\delta_s$ is no longer proportional to the field (see Fig. 2) and a beat structure develops with the nodes or so called "spin-zeroes" appearing at definite field values.

Because of the mixing between the Landau levels the level crossings are suppressed leading to so-called anti-crossing effect which is evident in Fig. 2 for exact solutions. Therefore, in general, it is no longer possible to identify a unique value of $\delta_s$ with each successive spin-zero, especially in the case of large spin orbit coupling. All solutions of the form $(n+1/2)\hbar\omega_c = \pm\delta_s$ (with *n* integer) are indistinguishable and additional analysis is required. For the exact (numerical) solutions in Fig. 2 the values of $\delta_s$ have therefore been chosen (somewhat arbitrarily) as the smallest positive solution of the matrix equation shown Fig. 1 (b), and all spin zeroes are found from $\delta_s = 1/2\,\hbar\omega_c$ and than are compared with WAL results as will be discussed later. For the approximate solution the dotted lines in Fig. 2 have been plotted as $2-\delta_s/\hbar\omega_c$ of the solution given by Eq. 2 (thin solid lines) so it matches the numerical solution for the second spin-zero corresponding to the spin sublevels crossing of the next upper Landau level.

It is clear from Fig. 2 that tilting the magnetic field has the strongest effect on the first spin zero and moves its position to higher magnetic fields (smaller values of $\Delta_R/\hbar\omega_c$, with $\Delta_R$ being fixed). This is illustrated in more detail in Fig. 3 where the position of the spin-zero is plotted against the tilt angle assuming $\Delta_R$ and $\hbar\omega_c$ being fixed for different values of parameter $\beta_z$, i.e. for different values of the effective *g*-factor. For comparison with the experimental data (see the discussion below) it is convenient to present these results by plotting appropriately normalized values of the square of the perpendicular component of magnetic field (at the spin-zero) against the square of the total field. As follows from Eq. (1) the position of the first spin-zero at zero tilt ($\cos\theta=1$) is given by:

$$\left(\frac{\Delta_R}{\hbar\omega_c}\right)^2_{\text{spin zero}} = \frac{9}{4} - (1-2\beta_z)^2 \qquad (3)$$

According to Eq. (3), for $\beta_z \ll 1$ the spin zero position ($\hbar\omega_c$ at the spin-zero position) depends essentially only on the value of $\Delta_R$, therefore for small values of the argument in Fig. 3 the curves all start (when the field is normal to the 2DEG plane) at $\left(\hbar\omega_c/\Delta_R\right)^2_{\text{spin zero}} = \frac{4}{5}$ or equivalently one can write $B_{\text{spin zero}} = 2\Delta_R m^*/\sqrt{5}e\hbar$.

## Experimental

Measurements were made in a $He_3$ cryostat, at temperatures of approximately 0.28 K. The magnetic field was tilted by rotating the sample in a split-coil, transverse axis super-conducting solenoid. A standard gated Hall bar was used (width 0.2mm) with measurements made, at 15 Hz, using standard AC techniques. The longitudinal and Hall resistances were measured simultaneously; this was found to be particularly important for the WAL feature near zero field because residual trapped flux in the split-coil magnet induced a slightly nonlinear and hysteretic relationship between magnet current and the magnetic field at the sample. Having the Hall voltage allowed accurate and reproducible determinations of both the tilt angle and the field at the sample.

Figure 4 shows the low field SdH oscillations at $V_g=0$ (corresponding to an electron density $n = 4.1\times10^{15}\text{m}^{-2}$). Results for several tilt angles plotted against $\rho_{xy} = B_\perp/en$ show, as expected, no change in the period of the oscillations with tilt. There is little evidence of a spin zero because its position lays in smaller field where SdH oscillations vanish. In this situation we see rather a general reduction of the amplitude over a wide field range expected for pure Zeeman splitting ($g^*\mu_B B \gg \Delta_R$) as mentioned above. This is because the data shown in Fig. 4 correspond to the situation of small $\Delta_R/\hbar\omega_c$ when the standard spin dependent term in the amplitude of the SdH oscillations, $\cos(\pi\delta_s / \hbar\omega_c)$, becomes important producing modulation of the SdH amplitude over a wide field range.[19, 20] Under these conditions the spin term reduces to $\cos(2\pi\beta_\perp / \cos\theta)$. As can be clearly seen in Fig. 4 for $\cos\theta$=0.083 there is a change in phase of 180° and the amplitude increase when $\beta_\perp/\cos\theta$ exceeds ¼ . Such phase changes can sometimes be observed several times when tilt is increased and indeed this is a well established technique for determining the electronic *g*-factor.[20]

Figure 5 shows the SdH amplitude of several peaks at three different fields (solid point) and fits to $\cos(2\pi\beta_\perp / \cos\theta)$. The fact that each fit yields a very similar value of $\beta_\perp$ is consistent with $\Delta_R$ being small. The average value of $\beta_\perp$ = -0.0316,[(d)] with an effective mass of 0.044 $m_e$, gives $g^* = -2.9$.

For a slightly higher density ($4.7\times10^{15}$ $m^{-2}$), shown in Fig. 6, a weak spin zero starts to be seen in normal field geometry (cos*θ*=1) at about 0.23 Tesla. In accordance with the theoretical prediction the beat structure moves to higher fields and becomes much more prominent when the field is tilted (cos*θ*=0.134). It is convenient to display the tilt dependence of the spin-zero by plotting the square of the perpendicular component of field at the spin zero against the square of the total field. This is shown in Fig.7 with the points as experimental data and the solid line a fit to Eq.(2) with β=-0.032. For reasons that are not fully understood an empirical linear fit actually provides a slightly better description of the data as shown by dashed line. This discrepancy is tentatively attributed to either a small change in effective *g*-factor with tilt and/or a small admixture of the Dresselhaus term in the predominantly Rashba spin-orbit interaction. It should be noted that the value of $\beta_\perp$ = -0.032 determined from the fit is in good agreement with the value obtained from the analysis of Figs. 4 and 5 when the spin-zero due to Rashba splitting was negligible.

According to theory the value of $\Delta_R$ in Fig. 7 is given, essentially, by the intercept of the extrapolated line at $B_{tot}$= 0 and is only weakly dependent on the value of β.[e] This means, that despite the small discrepancy between the expected and the observed dependences shown in Fig. 7, an empirical extrapolation of the tilted field data back to zero tilt is sufficient to provide an accurate estimate of $\Delta_R$ . Note also the fact that the spin-zero is moved to higher fields with tilt confirms that the beat can be correctly associated with spin-orbit splitting and is not due to some spurious artifact such as an inhomogeneous electron density, parallel conduction or conduction due to the second subband being occupied.

Because of the enhancement the position of the spin-zero in tilted fields can be followed over a range of densities. Results obtained in this way, at a fixed tilt of cos*θ*=0.174, are shown in Fig. 8. Here, to better define the position of the spin-zero, the second derivative of the transverse magnetoresistance was used, plotted as a function of total field for different gate voltages. The values of $\Delta_R$ obtained after taking into account the correction due to the tilt dependence (as in Fig. 7) are shown in Fig. 9 as solid circles, expressed in terms of the coefficient $\alpha$ given by $\Delta_R=2\alpha k_F$ where $k_F = \sqrt{2\pi n}$ is the electron Fermi wave vector. The open squares are results obtained from WAL measurements that are discussed in the next section.

---

[e] Using β= +0.032 rather than -0.032 would increase the value of $\Delta_R$ only by about 10%.

## Weak antilocalization measurements

As discussed above the general decrease of electron mobility with decreasing density means that the spin zeroes are often not visible, especially at low concentration (i.e. for negative gate voltages on sample considered here). The only way to then determine the strength of the spin-orbit interaction with a transport measurement is to study the WAL dip in resistivity around B=0. The WAL feature is more prominent in lower mobility samples but, with care, can usually be observed in any sample where it is present, i.e. when the spin-orbit interaction is sufficiently strong that the corresponding scattering time $\tau_{so}$ is shorter than the phase breaking time $\tau_\varphi$. To determine $\tau_{so}$ and $\tau_\varphi$ the experimental curves are fitted by an appropriate theoretical expression. In metals only the bulk Dresselhaus, cubic in *k*-vector, term is dominant and the WAL effect is then well described by the Hikami, Larkin and Nagaoka (HLN) equation.[3] In semiconductors, when linear in *k* Rashba term dominates, the HLN theory fails and it is more appropriate to use the theory by Iordanskii *et al.*[4] with two fitting parameters, $\beta_{so}=\tau_{tr}/\tau_{so}$ and $\beta_\varphi=\tau_{tr}/\tau_\varphi$. An example of experimental data fitted using this theory, in another InGaAs/InP sample with smaller spin-orbit term[21], is shown in Fig. 10 (a). The results are plotted against $B/B_{tr}$, where $B_{tr} = \hbar/4eD\tau_{tr}$ with $D=l^2/2\tau$ the diffusion constant, $l$ the mean free path, and $\tau$ the elastic momentum scattering time. It is evident from Fig. 10 (a) that in the case of small spin orbit coupling a good fit can be obtained both in the low field (negative magnetoconductance) region where the behavior is dominated by $\beta_\varphi$ and also for the higher field tails defined largely by spin-orbit parameter $\beta_{so}$[8]. This occurs because in this sample the conditions, $\beta_{so}<<1$ and $B/B_{tr}<<1$, for the Iordanskii *et al.* theory [4] to be valid are fulfilled. Deviations from the fit are noticeable as expected for the high field region ($B/B_{tr} \geq 1$).

Figure 10 (b) shows an attempt to make a similar fit for the sample under the study here, for which spin-zero data has been presented. In this case it is not possible to fit the data. The main reason for this is the larger spin-orbit coupling in this sample compared with the sample shown in Fig. 10 (a). That the spin orbit coupling is strong is qualitatively seen from the data presented in Fig. 10 (b) where the WAL feature extends to fields much larger than one, up to $5B_{tr}$, well beyond the range for which the theory in ref. 4 is valid. The turn over from WAL to weak localization effect also occur at high values of $B/B_{tr}>1$. The fits shown in Fig. 10 (b) have been made to the low field region ($B < B_{tr}$) where the theory is most likely still correct and where it probably gives a reasonable value for $\tau_\varphi$. Under these conditions, however, it gives false value for $\tau_{so}$, except that $\tau_{so}<\tau_\varphi$. It is stressed again that the Iordanskii *et a*l. theory is expected to fail when either $B/B_{tr}$ or $\beta_{so}$ becomes larger than one. The existing theoretical approach suggested by Laynda-Geller[22] is derived in the limit of strong magnetic fields ($B>B_{so}$) and is not applicable for fitting in the whole magnetic field range.

A new theory has been recently proposed for arbitrary strong spin-orbit coupling that may be able to deal with this case[23] although it has not, as yet, been subjected to an independent test against experimental data. It should be noted that the equations are relatively complicated and the theory is not very amenable to any straightforward fitting procedure.

It appears therefore, that in this and similar samples, where spin zeroes are not visible but where the mobility is too large (in other terms $\tau_{tr}>\tau_{so}$) there is currently no reliable means to determine, quantitatively, the magnitude of the spin-orbit interaction. To deal with this problem we note that in Fig. 10 (a) the turn-over from the low field WAL region to the weak-localization behavior occurs at a point very close to the fit value of parameter $\beta_{so}$ (marked by arrows) corresponding to the characteristic field $B_{SO} = \beta_{SO} B_{tr} = \hbar/4eD\tau_{SO}$. Therefore, the strength of the spin-orbit interaction can be quickly evaluated using $B_{so} \approx B_{min}$. We have verified this statement by numerical simulations (see Fig. 10 (a)) that it works within an experimental (≤10 %) error for the observed phase relaxation times and for the spin-orbit strength varied from the smallest when the WAL peak almost vanishes up to large values where the Iordanski et al. theory is still applicable. It should be mentioned that this assumption has also been used for a qualitative analysis in ref. 24. We therefore make the assumption, which will be tested *empirically* below, that even though the parameter $\beta_{so}$ may have a value larger than unity and that the physical significance of $\tau_{so}$ might not be as straightforward, the turn-over in the magnetoresistance still occurs at a field $B_{so}$ that can be related in the standard way, using the Dyakonov-Perel mechanism, to the spin-orbit energy [6], i.e.:

$$B_{SO} = \frac{2\Delta_R^2 \tau_{tr}}{4eD\hbar} \qquad (5)$$

Results obtained in this way, from the position of the minima in the WAL curves at various densities, are shown in Fig. 9 by open squares. There is a gratifyingly good agreement between the WAL and the spin-zero results. The fact that the relevant $\Delta_R$ depends on the square root of $B_{so}$ reduces to some extent the error involved in using the described empirical approach.

The good agreement for densities where values can be obtained from both spin-zeroes and the WAL, and also the fact that the same linear dependence is observed, gives some confidence that the suggested empirical procedure is giving us reasonable results. The magnitude of the spin-orbit coupling coefficient depends on many variables, e.g. the structure and doping details, interface properties, carrier concentration, strain, gate voltage, etc.[7,21,24] This gives us the flexibility in engineering and controlling the spin-orbit properties. On the other hand, each structure should be characterized individually. Therefore it is difficult to compare different experiments made on different samples, and it was our goal to perform two different experiments on the same sample. Nevertheless our results are in a qualitative agreement with other authors. The values of the spin-orbit constant have the same order of magnitude with those obtained on similar structures; in particular our results in Fig. 9 are very close to that obtained by Koga *et al.*[7] on sample No. 3.

## Tilt dependence of the weak anti-localization peak

It is known that parallel (in plane) magnetic fields can affect the low field magnetoresistance. In InGaAs quantum well samples two effects have been reported: the Zeeman splitting can modify the spin dephasing term[25] and the interface roughness effect produces an effective perpendicular component to a nominally pure parallel field.[5] It has been demonstrated in Ref. 25 that the Zeeman splitting associated with parallel fields of order 0.3 Tesla is sufficient to suppress the WAL peak and that fields of order 1 Tesla decrease $\tau_\varphi$ by a factor of order 4. Over the same field range $\tau_{so}$ was observed to increase (by over 30% in some instances). Minkov et al.[5] argue that in their sample for parallel fields surface roughness produces an effect of the opposite sign which largely cancels the Zeeman effect. It is of interest to understand whether such effects might be important in our sample. On the other hand, Meijer *et al.* [25] did not observe the surface roughness effect in their samples, which might be due to better surface quality.

In Fig. 11, which shows the shape of the WAL for three tilt angles from the normal (0°, 72.5° and 84.6°), it is clear that under these experimental conditions there is no change in the shape of the peak and, correspondingly, no detectable change in the values of $\tau_\varphi$ and $\tau_{so}$. The maximum perpendicular component of field shown here is ~15 mT corresponding to a maximum in-plane field of only 160 mT. Also, because of the geometry used, the total field at the centre of the peak is always zero so it is not surprising that these experiments see no change in the low field, WAL, portion of the trace. Measurements of the spin zero in tilted fields, however, use larger values of parallel fields but in this case, because the quoted values of $\Delta_R$ always involve an empirical extrapolation back to the zero tilt, any spurious parallel field contribution has been removed. Furthermore, the spin zero technique measures $\Delta_R$ directly and does not involve the scattering and dephasing dynamics that cause changes in $\tau_{so}$.

Note that to measure accurately the effect of parallel fields requires a different geometry; two orthogonal coils able to separately generate a large in-plane field and a small perpendicular field. Preliminary experiments indicate, however, that in-plane fields of order 0.6 Tesla are indeed sufficient to suppress the WAL peak and that the behavior is at least qualitatively similar to that seen in ref. 25. A detailed analysis would be complicated by the fact that another type of behavior is observed here: an extra magnetoresistance, quadratic in field, weakly dependent on density which, as is shown in Fig. 12, changes by a factor of ~2 when the alignment between the field and the current in the Hall bar is moved from normal to parallel configuration. This is attributed to the distortion of the Fermi contour associated with strong parallel fields. In a theoretical treatment, Heisz and Zaremba[26] find the same anisotropic dependence on current direction and with a magnitude (for GaAs/AlGaAs heterojunctions) roughly a factor two larger. The presence of this effect will complicate any analysis aimed at determining quantitatively the role of Zeeman splitting and interface roughness in this sample. It does not affect the results presented here for the strength of the spin-orbit term.

## Conclusion

Results of magnetotransport measurements are presented for a gated InGaAs quantum well sample where the spin-orbit interaction is strong and the disorder relatively small. In this condition the spin zeroes and the weak antilocalization effect can be simultaneously observed in the same sample. It has been demonstrated, we believe for the first time, that in this case both techniques give the same value for strength of the Rashba spin-orbit coupling.

To make this comparison it was necessary to enhance the strength of the spin-zeroes by tilting the magnetic field and a simple analytical approximation is presented to describe the modification of the coupling between the Zeeman and spin orbit effects introduced by tilted magnetic field. The relatively high mobility complicates the analysis of the weak antilocalization peak, because the transport time becomes longer than the spin-orbit scattering time, and an empirical procedure is proposed to deal with this situation, which is proved to work by comparing with the spin-zero results.

## Acknowledgements

We thank Jean Lapointe and other members of the Microfabrication Group for making gated Hall bar devices. GY was partially supported by DARPA-QUIST program.

**Electron spin-orbit splitting in InGaAs/InP quantum well studied by means of the weak antilocalization and spin-zero effects in tilted magnetic fields.**

S. A. Studenikin, P. T. Coleridge, G. Yu, and P. J. Poole

# Figures

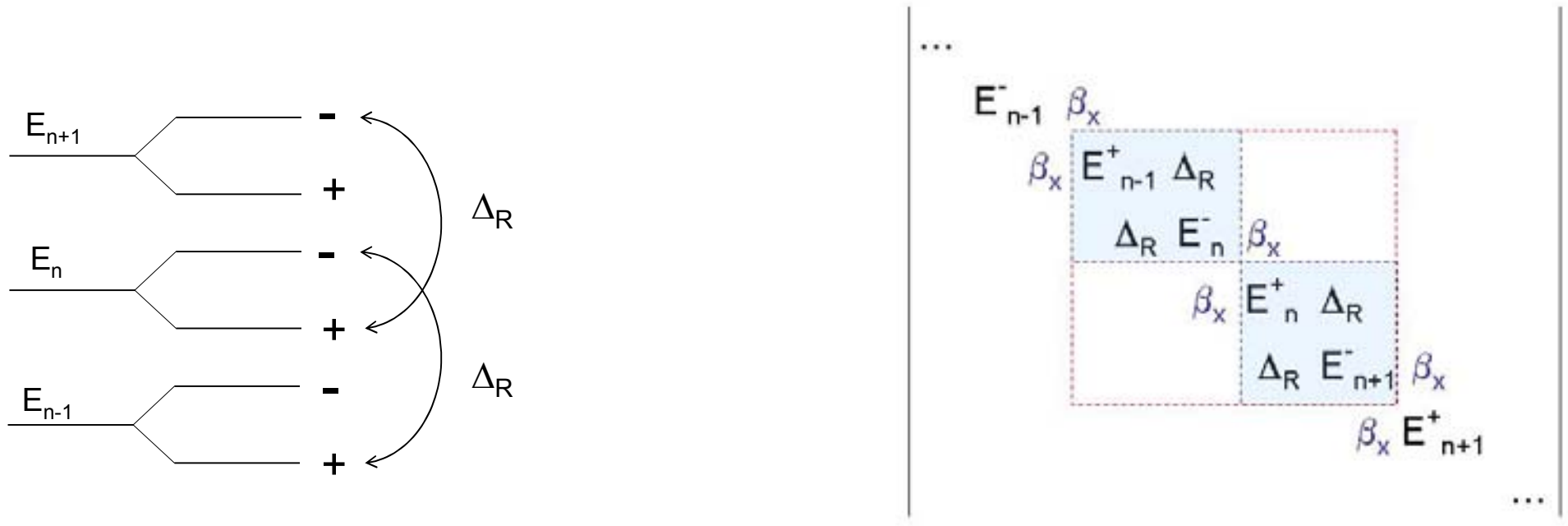

Figure 1. (a) Schematic presentation of spin-split Landau levels $E_n$. Spin orbit term $\Delta_R$ couples spin states of different Landau levels (reproduced from Das et al. [10]);
(b) matrix presentation of the Hamiltonian to determine the electron energy spectrum in tilted magnetic field in presence of the Zeeman and spin orbit terms. Diagonal terms, $E^{\pm}_n$, also include the perpendicular part of the Zeeman splitting $\pm\beta_z$.

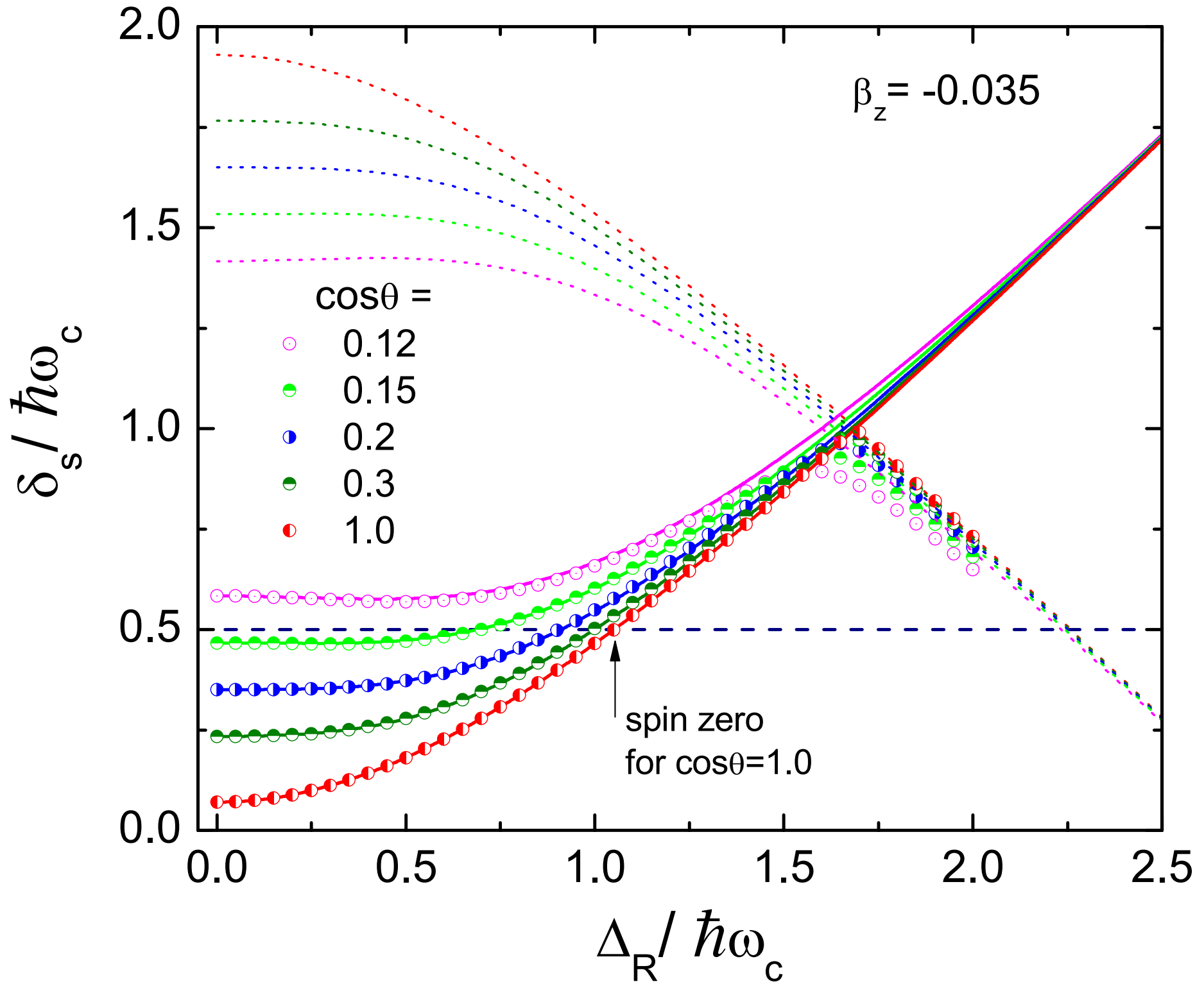

Figure 2. Calculated total spin-split for different tilts as a function of spin-orbit term in normalized scales. Points are exact solution of the Hamiltonian based on the tri-diagonal matrix in Fig. 1 (b). Solid lines are approximate analytical solution given by Eq. (2). Dashed declining lines are due to the spin term of the next higher Landau level at different tilts, same as for solid lines.

First spin zero position for cosθ=1 is marked by arrow. Second spin zero corresponds to the situation when the spin splitting is one and a half times bigger that the cyclotron energy and would occur at $\Delta_R/\hbar\omega_c \approx 2.2$.

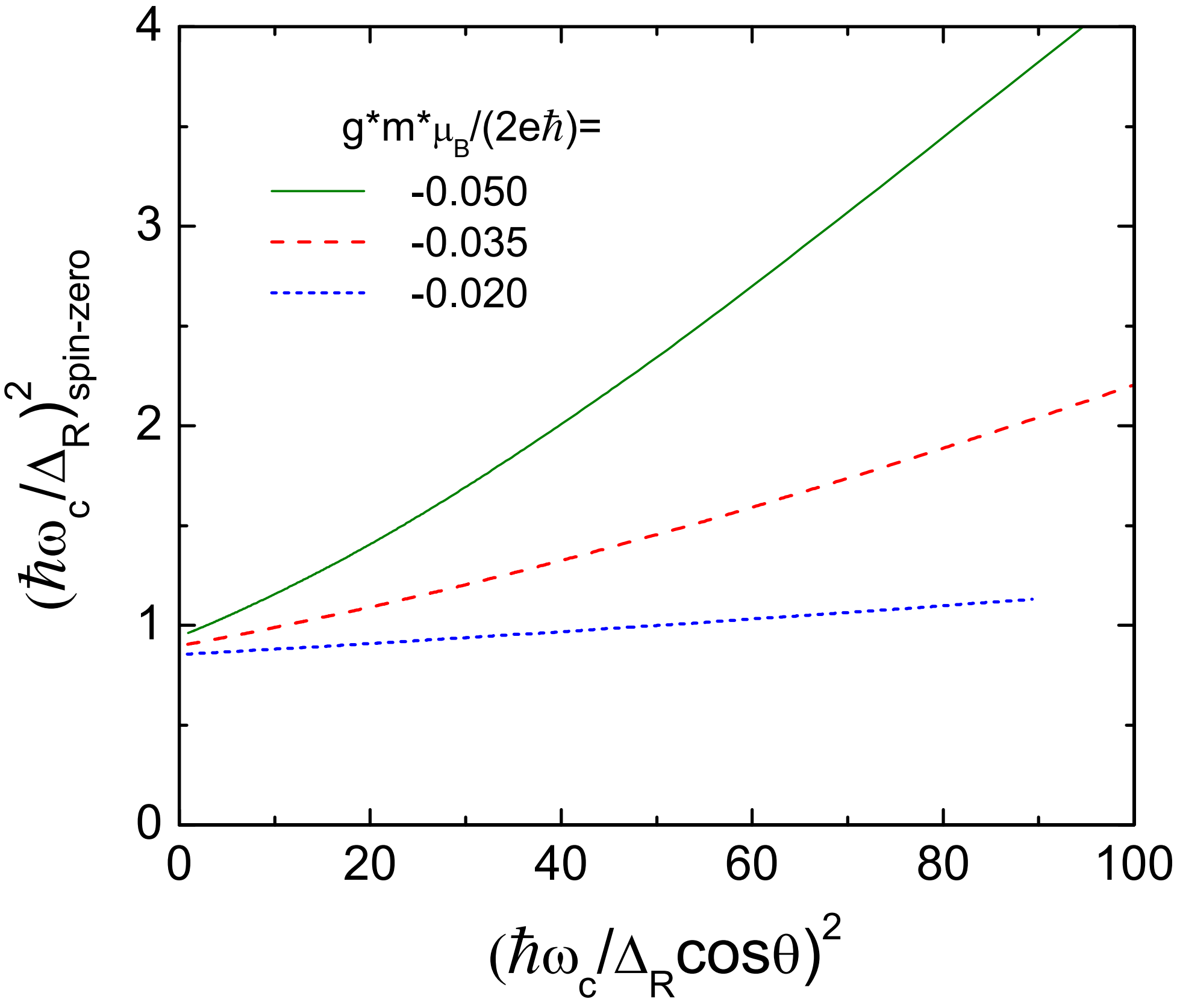

Figure 3. Position of the first spin zero as a function of the tilt angle ($\Delta_R$ and $\hbar\omega_c$ are assumed to be fixed during the tilting) for different values of parameter $\beta_z = g^* m^* \mu_B/(2e\hbar)$.

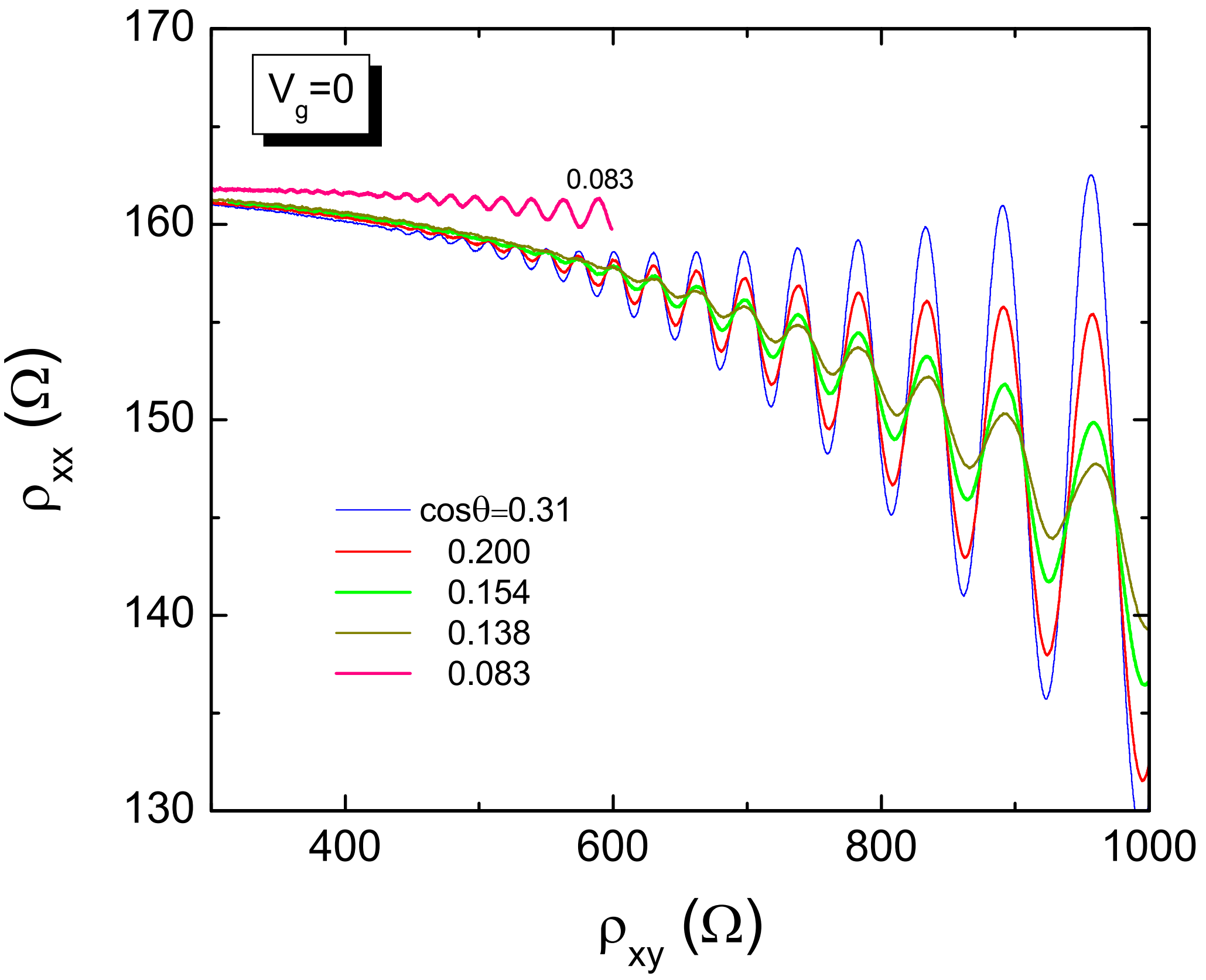

Figure 4. Low field SdH oscillations at $V_g$=0 ($n = 4.1\times10^{15}m^{-2}$) for different tilt angles from the normal to the surface. Results are plotted against $\rho_{xy} = B_\perp / en$ with 1000 Ω corresponding to normal component of the field $B_\perp$=0.65 T.

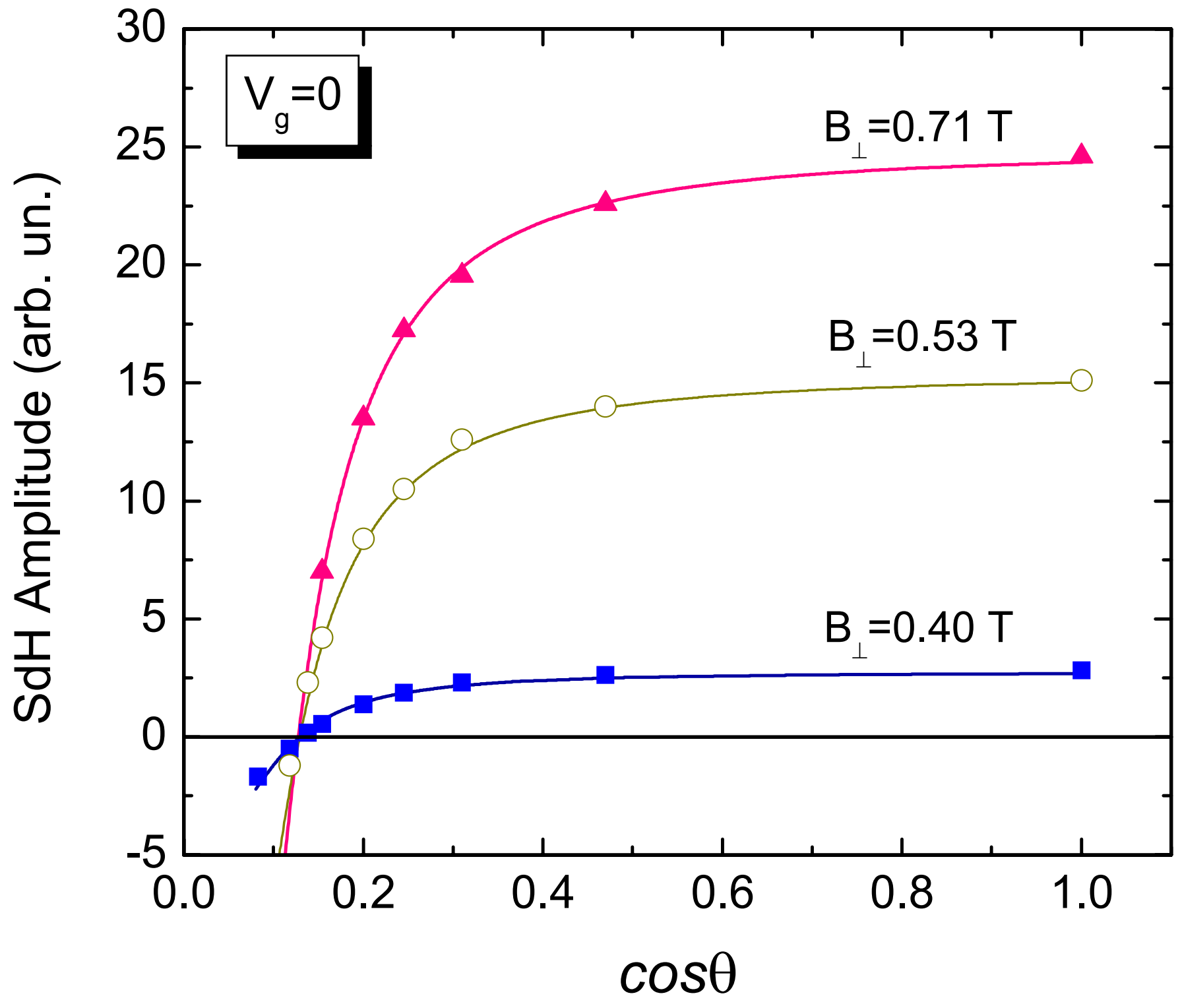

Figure 5. Points are the amplitude of the SdH oscillations at three different normal fields $B_\perp$ (filling factors ν=24, 32 and 42) as a function of the tilt angle.
Experimental data are fitted by function $A\cos(2\pi\beta_\perp/\cos\theta)$ with the following best values of $\beta_\perp$ 0.0320, 0.0312, and 0.0316 for magnetic fields 0.4, 0.53 and 0.71 T correspondingly.

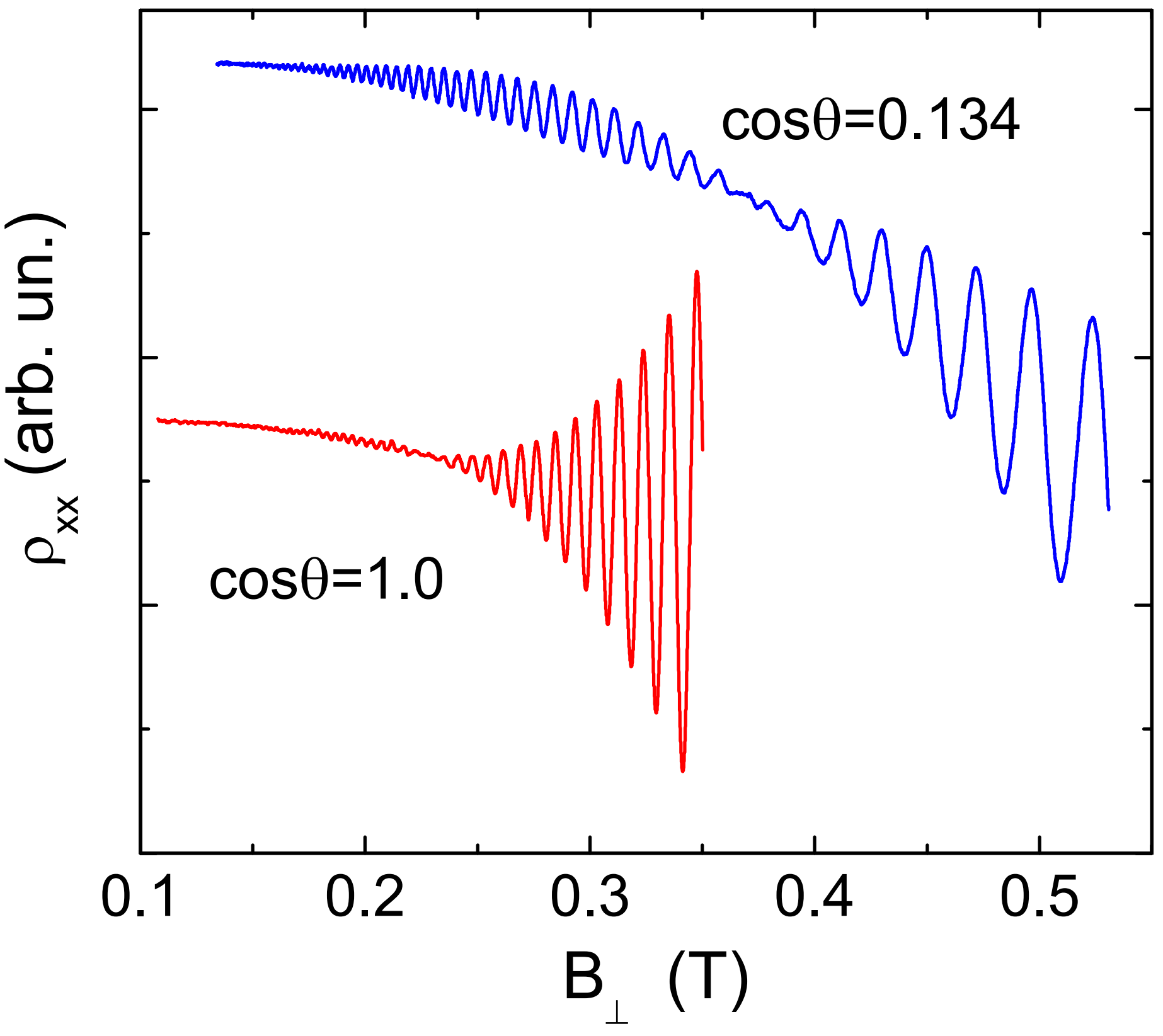

Figure 6. Tilt dependence of the spin zero in SdH oscillations for electron concentration of $4.7\times10^{15}$ $m^{-2}$ ($V_g$=0.32 V). The beat moves to higher fields and becomes much more prominent when the field is tilted (curve cosθ=0.134).

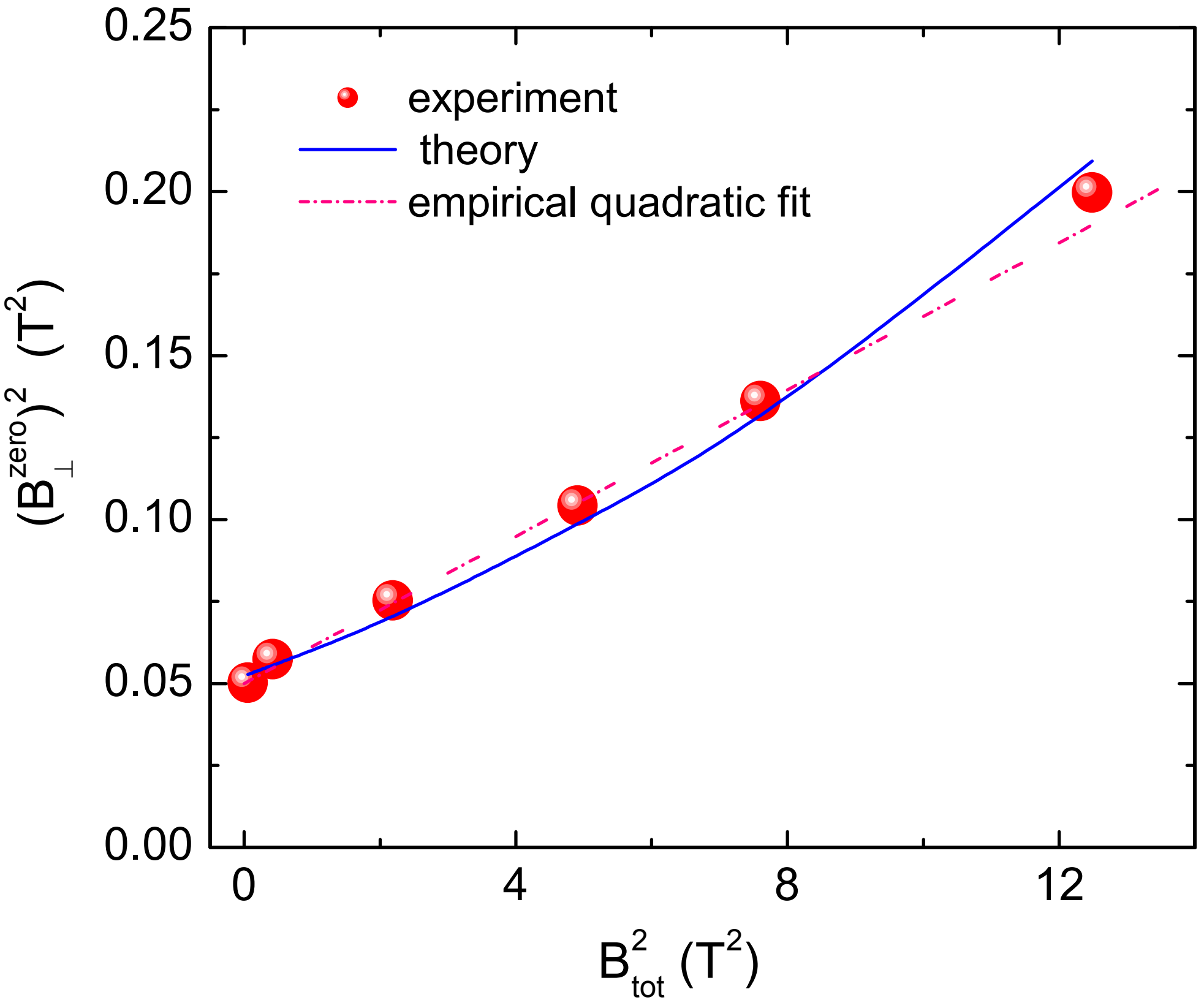

Figure 7. The square of the perpendicular component of magnetic field at the spin-zero against the square of the total field. The data were obtained from the SdH traces as shown in Fig. 6 at different tilt angles. Points are the experimental data, solid line is theoretical fit from Eq. (2), and dashed line is empirical linear fit in the quadratic scales.

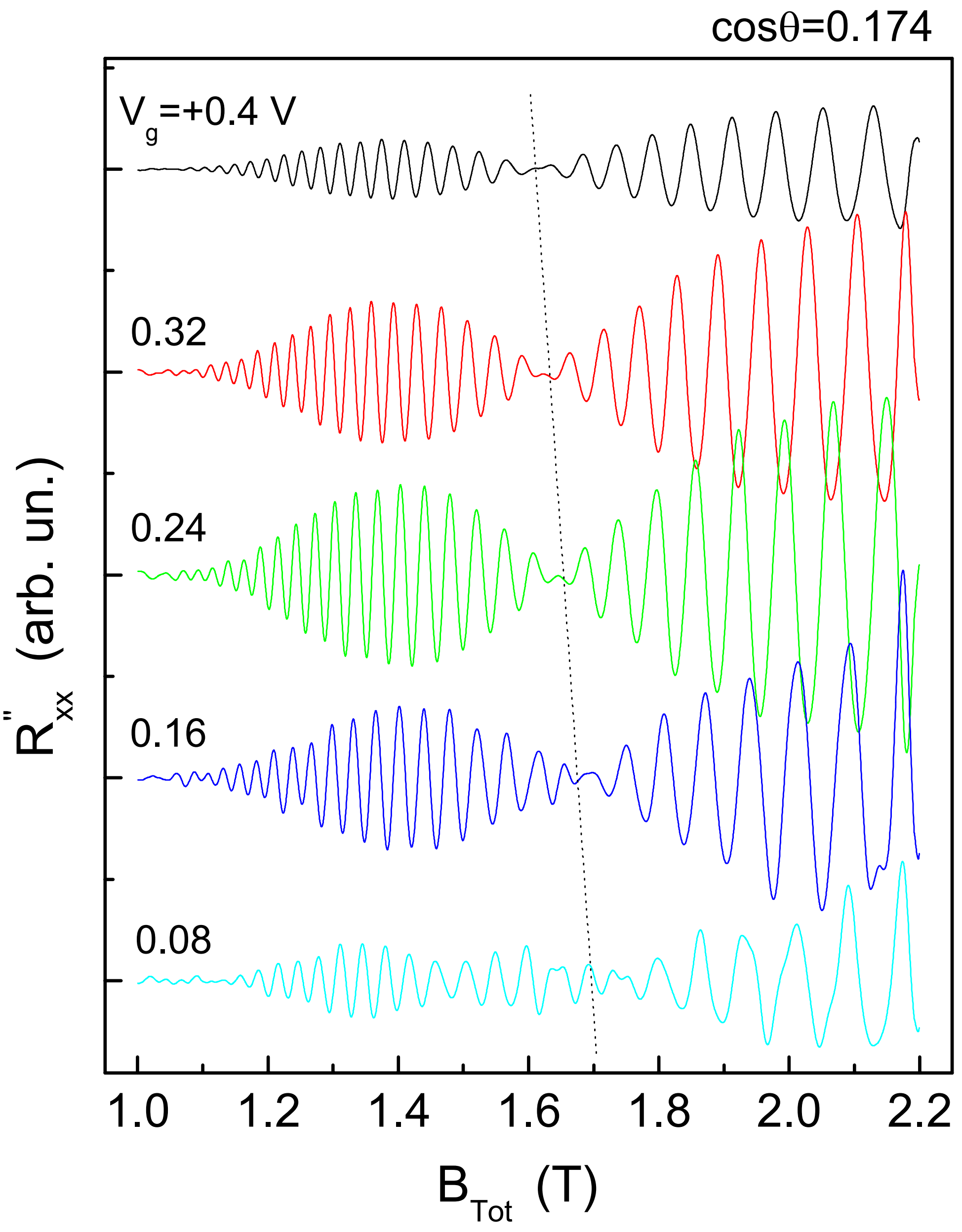

Figure 8. An example of the spin-zero position for different gate voltages at cosθ=0.174 tilt as a function of the total filed. For clarity and better precision of the beat position, a second derivative of the SdH oscillations is plotted and curves are shifted vertically.

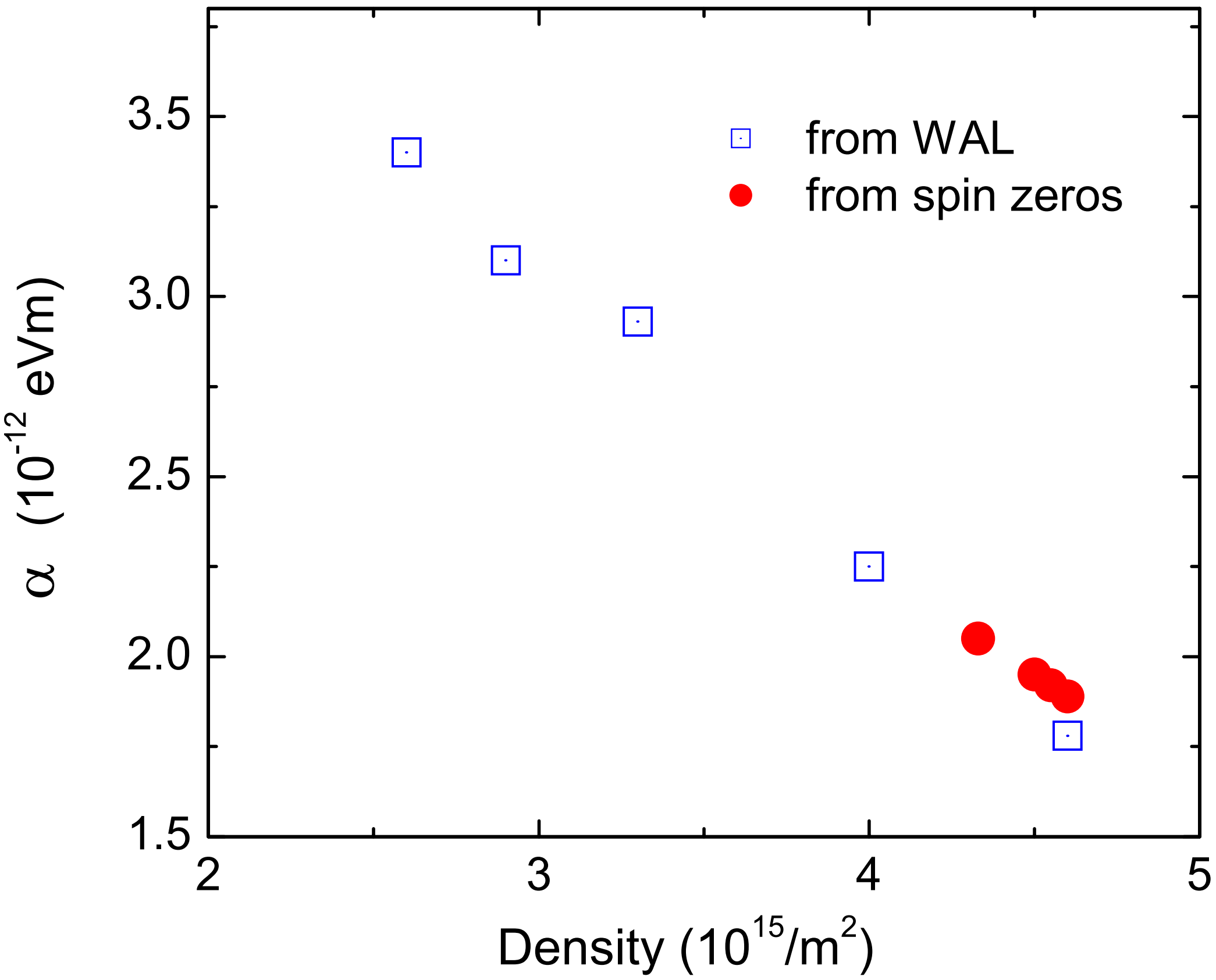

Figure 9. Spin orbit coefficient determined from spin zero beats in the SdH oscillations (solid circles) and from the weak antilocalization effect (open squares).

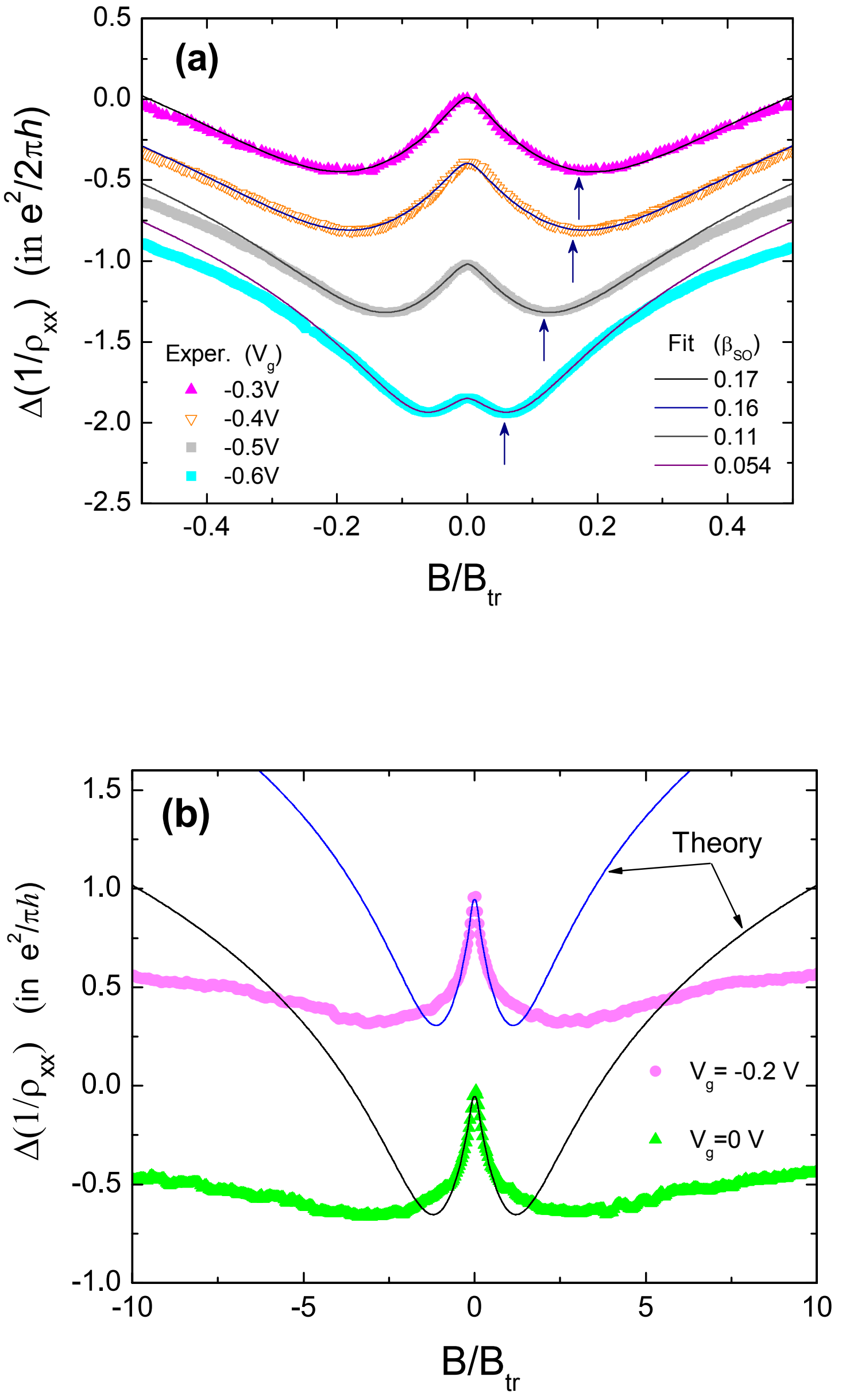

Figure 10. Weak antilocalization effect (points) fitted with theory Ref. 4 (solid lines).

(a) sample with smaller spin-orbit term studied earlier in ref. 21. The arrows are placed at the fitted values of $\beta_{so}$.

(b) sample studied in this work, which has large spin-orbit term resulted in the overturn from negative to positive magneto-conductance occurs at large, more than one, values of $B/B_{tr}$ when theory by Iordanski et al.[3] fails.

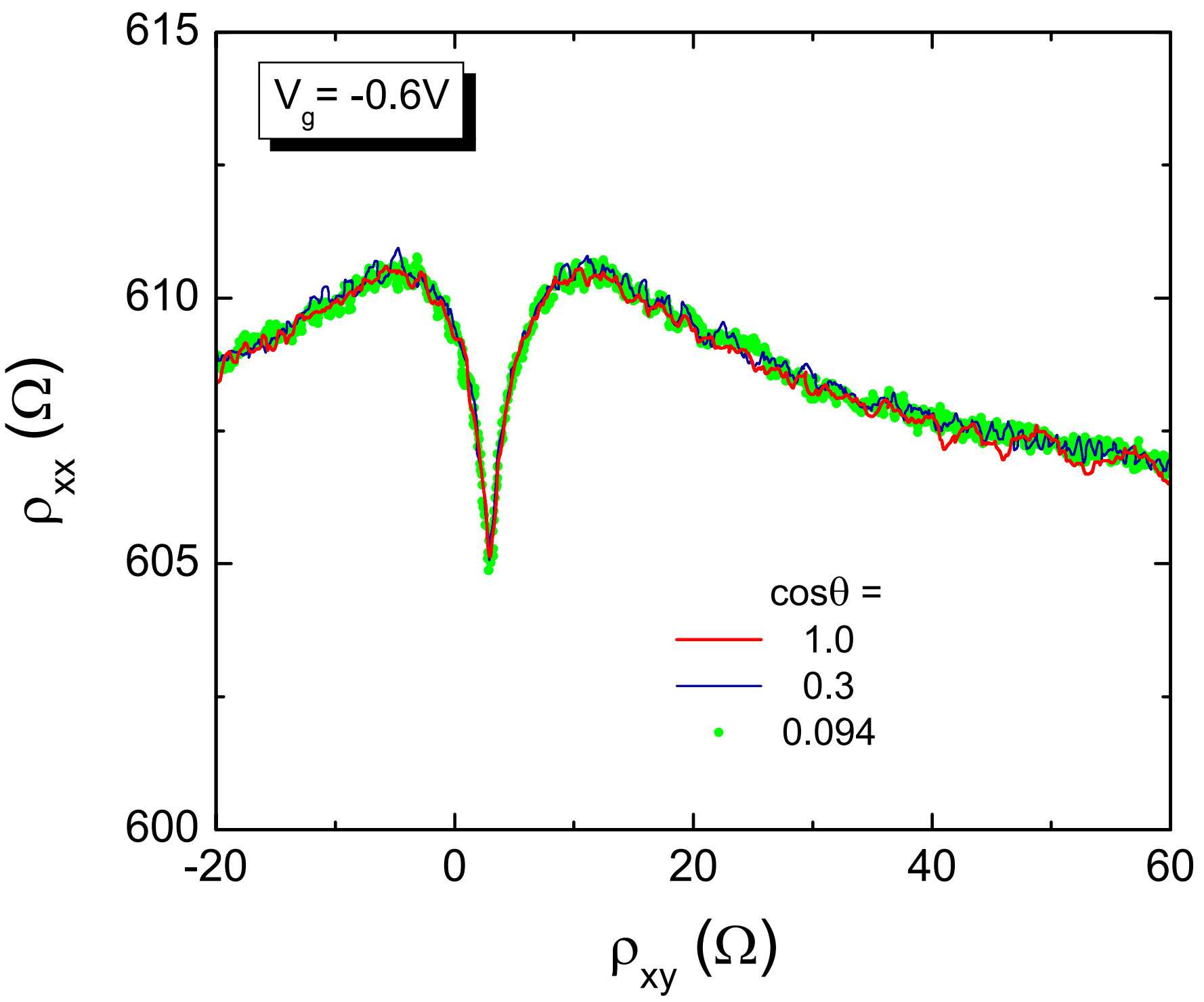

Figure 11. Magnetoresistance (WAL effect) plots for different tilts of the magnetic field plotted as a function of the Hall component $\rho_{xy}$=B/en, with $n$=1.8×10$^{15}$m$^{-2}$ and $\rho_{xy}$=3.50Ω at $B_{\perp}$=1mT.

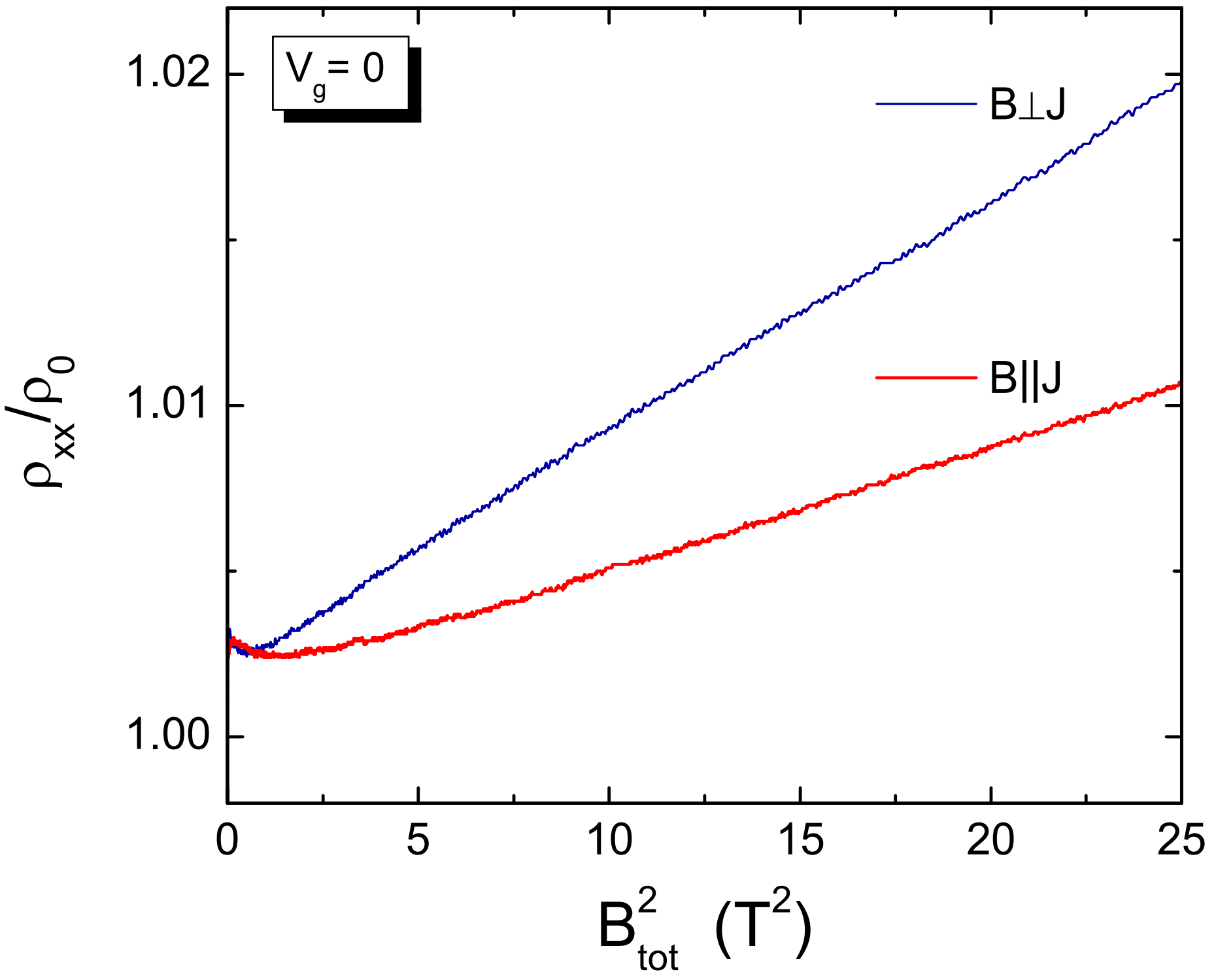

Figure 12. An example of the positive magnetoresistance caused by strong, purely in-plane field in two geometries when field is directed along and perpendicular to the current.